\documentclass[aps,prl,preprint,groupedaddress,amsmath,amssymb]{revtex4-1}

\usepackage{graphicx}
\usepackage{dcolumn}
\usepackage{bm}

\begin{document}




\title{Phase-stable limited relativistic acceleration or unlimited relativistic acceleration in the laser-thin-foil interactions}

\author{Yongsheng Huang}
 \homepage{http://www.anianet.com/adward}
\email{huangyongs@gmail.com}
\affiliation{China Institute of Atomic Energy, Beijing 102413, China.}%
\author{Naiyan Wang }
\author{Xiuzhang Tang }
\affiliation{China Institute of Atomic Energy, Beijing 102413,
China.}

\author{Yan Xueqing}
\affiliation{State Key Laboratory of Nuclear Physics and Technology,
Institute of Heavy Ion Physics, Peking University, Beijing 100871,
China}


\date{\today}

\begin{abstract}
To clarify the relationship between phase-stable acceleration (PSA)
and unlimited relativistic acceleration (URA) (Phys. Rev. Lett. 104,
135003 (2010)), an analytical relativistic model is proposed in the
interactions of the ultra-intense laser and nanometer foils, based
on hydrodynamic equations. The dependence of the ion momentum on
time is consistent with the previous results and checked by PIC
simulations. Depending on the initial ion momentum, relativistic RPA
contains two acceleration processes: phase-stable limited
relativistic acceleration (PS-LRA) and URA. In PS-LRA, the potential
is a deep well trapping the ions. The ion front, i.e., the bottom,
separates it into two parts: the left half region is PSA region; the
right half region is PSD region, where the ions climb up and are
decelerated to return back. In PS-LRA, the maximum ion energy is
limited. If the initial ion momentum large enough, ions will
experience a potential downhill and drop into a bottomless abyss,
which is called phase-lock-like position. URA is not phase-stable
any more. At the phase-lock-like position the ions can obtain
unlimited energy gain and the ion density is non-zero. You cannot
have both PSA and URA.
\end{abstract}


\maketitle

Laser-ion acceleration has been an international research
focus\cite{MakoTajima,Machnisms,Esirkepov,Yin}, however it is still
a challenge to obtain mono-energetic proton beams larger than
$100\mathrm{MeV}$. As a promising method to generate relativistic
mono-energetic protons, radiation pressure acceleration (RPA) has
attracted more attention\cite{Esirkepov,Yin,EsirkepovPRL96,Henig}
and becomes dominant in the interaction of the ultra-intense laser
pulse with nanometer foils. The phase-stable
acceleration\cite{YanxqPRL} predicted that the energy spread can be
improved in the interactions of nanometer-foils with
circular-polarized laser pulses. With thin-shell
model\cite{unlimitedRPA}, Bulanov and coworkers\cite{unlimitedRPA}
pointed out that the ions can obtain unlimited energy gain by RPA in
the relativistic limit. Yan and coworkers tried to predict the ion
energy distribution with a self-similar hydrodynamic
theory\cite{Yanxq}. However, it is nonrelativistic and under the
plasma approximation which allows $\nabla \bullet E\neq0$ and
$n_i-n_e=0$ satisfy together, where $E$ is the acceleration field
and $n_i$ (or $n_e$) is the ion (or electron) density. However, can
the ions obtain unlimited energy gain in the phase-stable region or
is the phase-stable acceleration still possible in the relativistic
limit? How about the relationship between them or what are the
critical conditions for them? To give clear answers of the
questions, an analytic self-consistent relativistic fluid model is
proposed to describe the relativistic radiation pressure
acceleration and to recheck the unlimited ion-acceleration region.

The ion acceleration in the interaction of ultra-intense laser and
nanometer foils contains two stages: the hole-boring process and the
radiation pressure acceleration. Here the transition of them is
assumed steady and the instability of the acceleration sheath is
suppressed well, which can be realized for specially designed
targets \cite{stableRPA1,stableRPA}. In the hole-boring process, the
ion velocity can reach $u_{hb}$\cite{Qiao2009,breaktime}, which is
the hole-boring velocity and also the initial velocity of the ions
in the radiation pressure acceleration. The initial time, $t_0$ is
when the compression layer is detached from the foil. It is decided
by the target thickness and $u_{hb}$. With the initial conditions:
$u_{hb}$ and $t_0$, the dependence of the ion energy on time can be
obtained and consists with the results of thin-shell
model\cite{Esirkepov} and has been checked by PIC
simulations\cite{Esirkepov}. Depending on the laser intensity, the
initial ion momentum will determine two different acceleration
processes: the phase-stable limited relativistic acceleration
(PS-LRA) and the unlimited relativistic acceleration (URA). When the
initial ion momentum is smaller than the critical one, the potential
is a deep well trapping the ions. The acceleration mode is PS-LRA
and contains the phase-stable acceleration region (PSA) and the
phase-stable deceleration region (PSD). The well is separated into
two half-regions: the left half-region and the right half-region by
the bottom, which is the ion front. The left half-region is PSA
region, where the electric field is positive and ions coast down to
the bottom. While the right half-region is PSD region, where the
electric field is negative and ions climb up the potential uphill
and are decelerated to return to the bottom. The deceleration is
also phase-stable.  No matter in PSA or PSD region, the maximum ion
energy is limited and ascertained by an analytical formulation.
Since PSA and PSD are separated by the ion front where the ion
density is zero, the ions in two regions can not exchange from each
other. If the initial ion momentum is large enough, the ions can get
across the potential uphill and experience a potential downhill and
then drop into the bottomless abyss which is the phase-lock-like
position. If the ions can reach the phase-lock-like position, they
can obtain unlimited energy gain as Bulanov and coworkers pointed
out\cite{unlimitedRPA}. The acceleration mode is URA and not
phase-stable any more. The electron density is smaller than the ion
density and the electron front increases with time. The
phase-lock-like position of URA is the limiting ion front and the
ion density is non-zero as time tends to zero. You cannot have both
PSA and URA in the same acceleration process. Therefore, the maximum
ion energy is finite in the phase-stable region and URA is not
phase-stable any more. However, no matter in PS-LRA or URA, the
plasma tends to neutral as time tends to infinite.


For convenience, the physical parameters: the time, $t$, the ion
position, $x$, the ion velocity, $v$, the electron field, $E$, the
electric potential, $\varphi$, the plasma density, $n$, and the
light speed, $c$, are normalized as follows: ${\tau}={\omega t},
\hat{x}=xk, u=v/c, \hat{E}={E}/{E_0}, {\phi}=\varphi/\varphi_0,
\hat{n}={n}/{n_{0}}, $ where $n$ represents $n_i$ (or $n_e$),
$n_{0}$ is the reference density, $\omega$ is the light frequency,
$k=\omega/c$ is the wave number, $c$ is the light speed,
$E_0={k\varphi_0}$, $e\varphi_0=\gamma_{em}m_ec^2$ and $\gamma_{em}$
is the maximum electron energy. Here $e$ is the elemental charge.

With reference to the results given by Mako and Tajima in Ref.
\cite{MakoTajima}, in the self-similar state, the density
distribution of ions is assumed as:
\begin{equation}\label{eq:ni}
\hat{n}_k=\frac{1}{\Sigma Q_k}(1+\phi)^{\alpha},k=1,...,N,
\end{equation}
where the subscribe $_k$ stands for the ion species, $Q_k$ is the
charge number of the $k$th species ion, the index $\alpha$ depends
on the laser intensity and the target thickness and discussed in the
Ref. \cite{Yanxq}.

With the transformation: $\xi=\hat{x}/\tau$, the normalized
continuity and motion equation of ions and Poisson's equation are
given as:
\begin{equation}\label{eq:ionsys}
\begin{aligned}
\left(u_k-\xi\right)\frac{\partial\ln
\hat{n}_k}{\partial\xi}=-\frac{\partial
u_k}{\partial\xi},\\
\left(u_k-\xi\right)\frac{\partial\gamma_k
u_k}{\partial\xi}=-\beta_k\frac{\partial \phi}{\partial\xi},
\end{aligned}
\end{equation}
\begin{equation}\label{eq:poi}
\frac{1}{\tau^2}\frac{\partial^2\phi}{\partial\xi^2}=-\rho
\left(\Sigma Q_k\hat{n}_k-\hat{n}_e\right)
\end{equation}
 where $\beta_k=\frac{Q_k\gamma_{em}m_e}{M_k}$, $M_k$ is the mass of
certain ions, $\gamma_k=(1-u_k^2)^{-1/2}$,
$\rho=\frac{\omega_{pe}^2}{\gamma_{em}\omega^2}$, and
$\omega_{pe}^2=\frac{n_0e^2}{\epsilon_0m_e}$.

Solving Eq. (\ref{eq:ionsys}), the ion velocity satisfies:
\begin{equation}\label{eq:ui}
\xi+\left(\frac{\gamma_k}{\gamma_{k,0}}\right)^{-3/2}u_{k,0}=u_k+\frac{\gamma_k^{-3/2}}{2\alpha}\left(\chi-\chi_0\right),
\end{equation}
and the potential in the ion region is given by:
\begin{equation}\label{eq:phi}
\phi_1=\frac{\left(\chi-\chi_0-2\alpha
u_{k,0}\gamma_{k,0}^{3/2}\right)^2}{4\alpha\beta_k}-1,
\end{equation}
where $\gamma_{k,0}=(1-u_{k,0}^2)^{-1/2}$, $\chi_0=\chi(u_{k,0})$,
\begin{equation}\label{eq:chi}
\chi=\int^{u_k}_0\gamma_k^{3/2}du_k,
\end{equation}
and $u_{k,0}$ is the hole-boring velocity given
by\cite{Qiao2009,breaktime}:
\begin{equation}\label{eq:uk0}
u_{k,0}=\frac{u_{hb}}{c}=\sqrt{\frac{Z}{A}\frac{m_e}{\gamma_{hb}M_k}\frac{n_c}{2n_0}}a
\end{equation}
where $a^2=0.732I_{10^{18}\mathrm{W/cm^2}}\lambda^2_{\mathrm{\mu
m}}$, $I_{10^{18}\mathrm{W/cm^2}}$ is the laser intensity in unit of
$10^{18}\mathrm{W/cm^2}$, and
$\gamma_{hb}={(1-(u_{hb}/c)^2)^{-1/2}}$. The beginning time $\tau_0$
is when the compressed ion and electron layer is detached from the
foil and given by\cite{breaktime}:
\begin{equation}\label{eq:bound}
\tau_0=\frac{d}{\lambda}\frac{2\pi }{u_{k,0}},
\end{equation}
at $\xi=0$.

\begin{figure}
{
\includegraphics[width=0.5\textwidth]{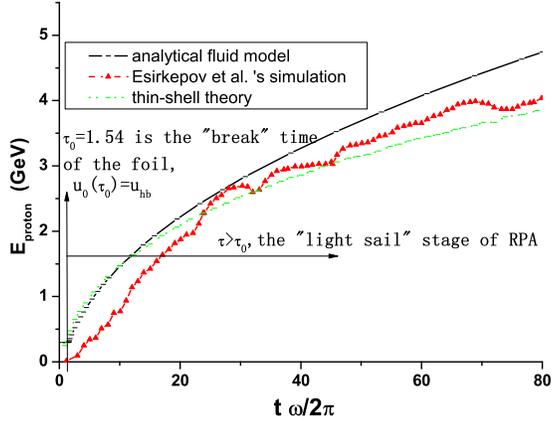}
}  \caption{\label{fig:Ecom0} (Color online) Comparison of our
analytical fluid model with thin-shell model and Esirkepov et al.'s
simulations for $\sigma/a\approx0.1$, $a=316$, $d=\lambda=1\mu m$,
$n_0=49n_c=5.5\times10^{22}/cm^3$, and $\alpha=1.8$\cite{Yanxq}.}
\end{figure}

In order to obtain the dependence of the ion velocity on time, it is
need to solve Eq. (\ref{eq:ui}). From Eq. (\ref{eq:ui}), the
following differential equations of two variables are given:
\begin{equation}\label{eq:pvk}
\left\{ \begin{aligned}
         \frac{dp_k}{dt}=\frac{2\alpha
V_k(1+p_k^2)^{3/2}}{t\left[3\alpha
p_kV_k\sqrt{1+p_k^2}-(2\alpha+1)t\right]},\\
\frac{dV_k}{dt}=\frac{-2\alpha V_k}{3\alpha
p_kV_k\sqrt{1+p_k^2}-(2\alpha+1)t}.
                          \end{aligned} \right.
\end{equation}
where $p_k=u_k\gamma_k$ is the normalized ion momentum and $V_k$
satisfies:
\begin{equation}\label{eq:Vk}
V_k=\int^t_0\frac{p_k}{\sqrt{1+p_k^2}}dt-t\frac{p_k}{\sqrt{1+p_k^2}}.
\end{equation}
Using matlab function $ode113$ to solve Eq. (\ref{eq:pvk}) with the
initial time $\tau_0$ and initial velocity, $u_{k,0}$, the
dependence of the ion energy on time has been calculated
\cite{matlabfile}. As an example, Figure \ref{fig:Ecom0} shows the
comparison of our analytical fluid model with the thin-shell model
and PIC simulations\cite{Esirkepov}. The results of our model
consist well with that of the PIC simulations. When $\tau$ is larger
than $40$ times of the laser cycle, our results are a little larger
than that of the PIC simulations. One of the reason is the loss of
laser energy and the decreasing of the electron temperature for
large $\tau$ in the simulations, while the electron temperature is
assumed to be a constant in our model.

Combing Eqs. (\ref{eq:ni}) and (\ref{eq:phi}), it is obtained:
\begin{equation}\label{eq:niuk}
\hat{n}_k=\frac{1}{\Sigma Q_k}\frac{\left(\chi-\chi_0-2\alpha
u_{k,0}\gamma_{k,0}^{3/2}\right)^{2\alpha}}{\left(4\alpha\beta_k\right)^{\alpha}},
\end{equation}

With Eqs. (\ref{eq:poi}), (\ref{eq:ui}) and (\ref{eq:phi}), the
electron density in the ion region is written as:
\begin{equation}\label{eq:ne}
\hat{n}_e=\left(1+\phi_1\right)^{\alpha}+\frac{1}{\rho\tau^2}\frac{\partial^2\phi_1}{\partial\xi^2},
\end{equation}
Equation (\ref{eq:ne}) shows the plasma can not be quasi-neutral at
a finite time. However, the plasma tends to neutral as the time
tends to infinite. It is obtained:
\begin{equation}\label{eq:neutral}
\lim_{\tau\rightarrow+\infty}n_e=\Sigma Q_kn_k,
\end{equation}

With Eq. (\ref{eq:niuk}) and $\hat{n}_k(\xi=\xi_{i,f})=0$, the
possible maximum ion energy at the ion front satisfies:
\begin{equation}\label{eq:uikm}
\int_0^{p_{k,m}}\gamma_k^{-\frac{3}{2}}dp_k=\int_0^{p_{k,0}}\gamma_k^{-\frac{3}{2}}dp_k+2\alpha
p_{k,0}\gamma_{k,0}^{\frac{1}{2}},
\end{equation}
where $\gamma_k^2=1+p_k^2$, $p_{k,m}=u_{k,m}/\sqrt{1-u_{k,m}^2}$ is
the normalized limiting momentum of ions at the ion front:
$\xi=\xi_{i,f}$ and $p_{k,0}=u_{k,0}/\sqrt{1-u_{k,0}^2}$.

Different from the nonrelativistic case, it contains two
acceleration modes and depends on the initial conditions: $u_{k,0}$
and $\alpha$ in the relativistic case. The critical condition is
decided by:
\begin{equation}\label{eq:pk0c}
\int_0^{p_{k,0}}\gamma_k^{-\frac{3}{2}}dp_k+2\alpha
p_{k,0}\gamma_{k,0}^{\frac{1}{2}}=\int_0^{+\infty}(1+x^2)^{-3/4}dx\approx2.622.
\end{equation}

\begin{figure}
{
\includegraphics[width=0.75\textwidth]{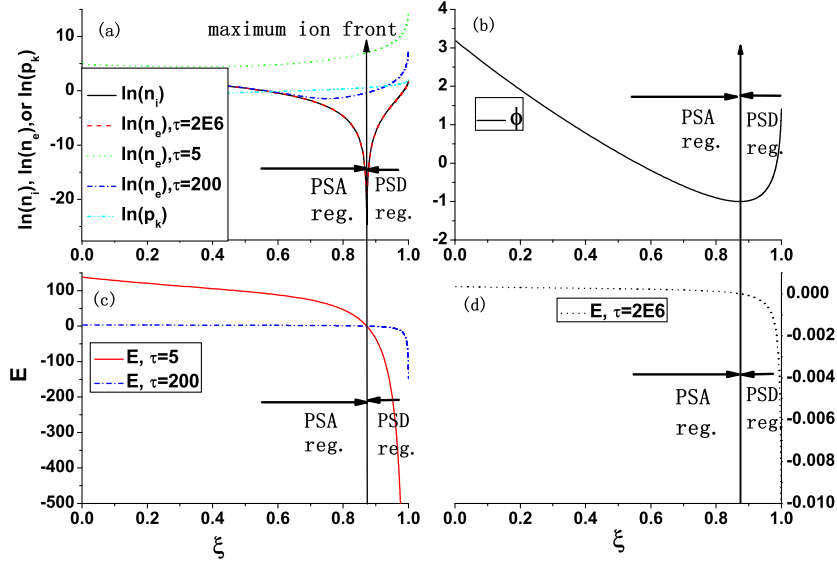}
}  \caption{\label{fig:ninelim} (Color online) Phase-stable limited
relativistic acceleration (PS-LRA) contains two regions:
phase-stable acceleration region for
$0\leq\xi\leq\xi_{i,f}\approx0.87$ and phase-stable deceleration
region for $0.87\leq\xi\leq1$. (a)The density of ions and electrons
for different time and the ion momentum VS the self-similar variable
$\xi$. At the ion front $\xi_{i,f}=0.87$, the ion density is zero.
(b)The potential VS $\xi$. (c) and (d) The electric field $\hat{E}$
VS $\xi$. $\hat{E}\geq0$ for $0\leq\xi\leq\xi_{i,f}$ and
$\hat{E}\leq0$ for $\xi_{i,f}\leq\xi\leq1$. Here,
$n_{0}=10^{22}\mathrm{/cm^3}$, $\alpha=2$ and $a=70$.}
\end{figure}

First, $\int_0^{p_{k,0}}\gamma_k^{-\frac{3}{2}}dp_k+2\alpha
p_{k,0}\gamma_{k,0}^{\frac{1}{2}}\lneq2.622$ for
$p_{k,0}\lneq0.5064$ and $\alpha=2$, which requires $a\lneq203$ for
$n_0=49n_c$, $d=\lambda=1\mu m$. Figure \ref{fig:ninelim} shows the
dependence of the plasma density, electric field and potential on
$\xi$ for $n_{0}=10^{22}\mathrm{/cm^3}$, $\alpha=2$ and $a=70$. In
this case, it is divided into two regions: the phase-stable
acceleration region for $0\leq\xi\leq\xi_{i,f}$ and the phase-stable
deceleration region $\xi_{i,f}\leq\xi\leq1$.

(I) In the phase-stable acceleration region,
$0\leq\xi\leq\xi_{i,f}$, the electric field $E\geq0$ and the
electron density is larger than the ion density. The maximum ion
momentum $p_{k,m1}$ is limited and given Eq. (\ref{eq:uikm}) at the
ion front $\xi_{i,f}$ ascertained by Eq. (\ref{eq:ui}) with
$u_k=u_{k,m1}$, where $u_{k,m1}=p_{k,m1}/\sqrt{1+p_{k,m1}^2}$. At
the ion front, the ion density is zero. The difference of the
electron density and ion density decreases with time and tends to
zero.

The potential shown by Figure \ref{fig:ninelim} (b) in
$0\leq\xi\leq\xi_{i,f}$ gives an intuitionistic explanation of PSA.
The ions coast down the slope of the potential, and the gradient,
i.e., the electric field, becomes gently as the ions come to the
bottom of the potential although few can reach there. Therefore the
ions at higher potential will obtain more acceleration and the
energy spread is improved. That is PSA. Different from the real
gliding process, the ions can not pass through the bottom and climb
up since the ion front is the limiting point and the ion density
tends to zero at the bottom.

(II) In the phase-stable deceleration region,
$\xi_{i,f}\leq\xi\leq1$, the eletric field $E\leq0$ and the electron
density is larger than the ion density too as shown in Figure
\ref{fig:ninelim}. In this region, the ion momentum $p_k$ satisfies:
$p_{k,m1}\leq p_k\leq p_{k,m2}$, where $p_{k,m2}=p_{k}(\xi=1)$ and
given by Eq. (\ref{eq:ui}) with $\xi=1$. All the ions in this region
are decelerated to $\xi=\xi_{i,f}$. The absolute value of the
electric field decreases with the decreasing $\xi$ and is zero at
the ion front. Therefore the deceleration is also phase-stable.

With Figure \ref{fig:ninelim} (b), for ions, the potential in
$\xi_{i,f}\leq\xi\leq1$ is a mountain with height of about
$35\mathrm{GeV}$ which is far larger than the maximum ion energy of
about $\sqrt{p_{m,2}^2+1}-1\approx 7\mathrm{GeV}$ at $\xi=1$.
Therefore, the ions will be decelerated at the potential uphill.
Since the gradient, i.e., the value of electric field increases with
the potential height, the deceleration is also phase-stable. As
point out above, the limiting point is still the ion front
$\xi=\xi_{i,f}$, and the ions in PSD region can also not pass
through the bottom into PSA region. The ions in PSA and PSD region
can not exchange from each other because of the zero-density
dividing point $\xi=\xi_{i,f}$.

In this case, the maximum ion momentum is $p_{k,m1}$ at the ion
front $\xi_{i,f}\lneq1$. Therefore, it is called phase-stable
limited relativistic acceleration (PS-LRA). The ions in the two
phase-stable regions can not exchange from each other and the ion
front $\xi_{i,f}$ is the dividing line. Combing the condition of
PS-LRA and the above discussion about PSA and PSD, the ions with an
initial momentum of $p_{k,0}$ not large enough to get across the
potential at $\xi=1$ will drop in PSA region or PSD region and
obtain a finite maximum energy ascertained by Eq. (\ref{eq:uikm}).

\begin{figure}
{
\includegraphics[width=0.75\textwidth]{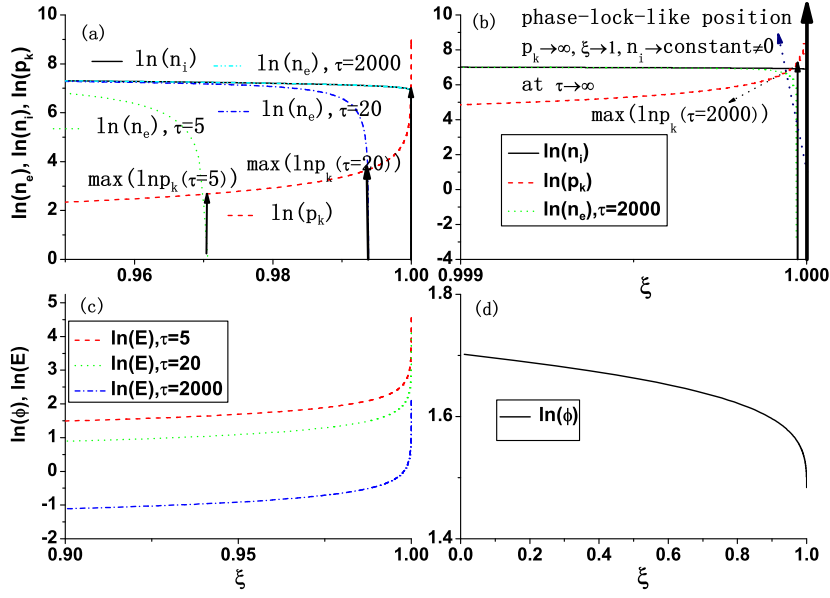}
}  \caption{\label{fig:nineunlim} (Color online) Unlimited
relativistic acceleration (URA) with the phase-lock-like position
$\xi=1$.(a)The density of ions and electrons for different time and
the ion momentum VS $\xi$. The ion density is no-zero at the
limiting ion front: $\xi=1$. The electron density is smaller than
that of ions at any finite time. (b)The enlargement of (a) near
$\xi=1$. $\xi=1$ is the phase-lock-like position where the ion
momentum tends to infinity. (c) and (d) The electric field $\hat{E}$
and the potential $\phi$ VS $\xi$. $\hat{E}\geq0$ for all the region
and the acceleration is not phase-stable any more. Here $a=316$,
$d=\lambda=1\mathrm{\mu m}$,
$n_0=49n_c=5.5\times10^{22}\mathrm{/cm^3}$, and $\alpha=2$.}
\end{figure}

If the initial ion momentum satisfies:
\begin{equation}\label{eq:pk0un}
\int_0^{p_{k,0}}\gamma_k^{-\frac{3}{2}}dp_k+2\alpha
p_{k,0}\gamma_{k,0}^{\frac{1}{2}}\geq2.622.
\end{equation}
the ions will not experience a potential well in PS-LRA and will
coast down from the potential slope and drop into the bottomless
abyss at $\xi=1$ as shown by Figure \ref{fig:nineunlim} (d). $\xi=1$
is called phase-lock-like position and the ions can obtain unlimited
energy gain as shown in Figure \ref{fig:nineunlim} (b). It is called
unlimited relativistic acceleration (URA), which is not phase-stable
any more. Eq. (\ref{eq:pk0un}) satisfies for $p_{k,0}\geq0.5064$,
i.e., $a\geq203$ from Eq. (\ref{eq:uk0}) for $\alpha=2$,
$n_0=49n_c=5.5\times10^{22}\mathrm{/cm^3}$ and
$d=\lambda=1\mathrm{\mu m}$. Figure \ref{fig:nineunlim} shows the
dependence of the plasma density, electric field and potential on
$\xi$ for $a=316$, $d=\lambda=1\mu m$,
$n_0=49n_c=5.5\times10^{22}/cm^3$, and $\alpha=2$. In this case, the
electron density is smaller than the ion density and the
acceleration is not phase-stable any more. In all the region, the
electric field increases with $\xi$ and is larger than zero. At a
finite time, the electron front, where $n_e=0$, $\xi_{e,f}\lneq1$,
which is 0.971, 0.994, 0.99997 at $\tau=5,20,2000$ separately.
Therefore the possible maximum the ion momentum is shown by Figure
\ref{fig:nineunlim} (a) and (b).

As discussed above, due to the initial ion momentum $p_{k,0}$ large
enough, the ions can reach URA region. Therefore the ion momentum
and the field have a sharp increase and tend to infinite, the ion
density becomes a non-zero constant in $\xi\in(1-\delta,1]$ as
$\tau\rightarrow\infty$, where $\delta$ is an infinitesimal. As
shown in Figure \ref{fig:nineunlim} (b), at the phase-lock-like
position and the limiting ion front $\xi=1$, the ion can obtain
unlimited energy gain and the ion density is non-zero. It is similar
with the unlimited phase-lock ion acceleration as pointed out by
Bulanov and coworkers\cite{unlimitedRPA} in the relativistic limit.
In URA region, it is found that (I) the unlimited ion acceleration
requires the initial ion momentum is large enough and should meet
Equation (\ref{eq:pk0un}); (II) it is not phase-stable any more;
(III) the phase-lock-like position is $\xi=1$, the limiting ion
front; (IV) the ion density at the limiting ion front is non-zero.

In the conclusion, it has been given an analytical relativistic
fluid model to describe the relativistic radiation pressure
acceleration with the initial parameters from the hole-boring stage.
The dependence of the ion velocity on the acceleration time can be
obtained and is consistent with that of thin-shell model and PIC
simulations. There are two acceleration modes: PS-LRA and URA with a
critical initial ion momentum ascertained by an explicit
formulation. In PS-LRA, the ions are trapped in a deep potential
well and the maximum ion energy is limited and the ion front is the
well bottom and $\xi_{i,f}\lneq1$. URA is not phase-stable any more
and there is a phase-lock-like position in it. At the
phase-lock-like position, corresponding to the relativistic limit,
the ions can obtain unlimited energy gain and the ion density is
non-zero as time tends infinite. Although the unlimited ion
acceleration can not be reached at any finite time, the ions can be
accelerated to any large energy if the laser pulse is long enough.
As an important result, you cannot obtain both PSA and URA.
Therefore, if the laser parameters are large enough to obtain URA,
the energy spread must be lost. If one wants to improve the energy
spread with PSA, the maximum ion energy is limited.

\begin{acknowledgments}
This work was supported by the Key Project of Chinese National
Programs for Fundamental Research (973 Program) under contract No.
$2011CB808104$ and the Chinese National Natural Science Foundation
under contract No. $10834008$.
\end{acknowledgments}


\begin{thebibliography}{19}

\bibitem{MakoTajima} F. Mako and T. Tajima, Phys. Fluids 27,
1815 (1984).

\bibitem{Machnisms}Y. Oishi, T. Nayuki, T. Fujii, Y. Takizawa, X. Wang, T. Yamazaki, K.
Nemoto, T. Kayoiji, T. Sekiya, K. Horioka, Y. Okano, Y. Hironaka, K.
G. Nakamura, K. Kondo, A. A. Andreev, Phys. Plasmas 12, 073102
(2005);H. Schwoerer, S. Pfotenhauer, O. Jackel, K.-U. Amthor, B.
Liesfeld, W. Ziegler, R. Sauerbrey, K. W. D. Ledingham, T.
Esirkepov, Nature 439, 445 (2006); M. Murakami and M. M. Basko,
Phys. Plasmas 13, 012105 (2006).

\bibitem{Esirkepov} T. Esirkepov, M. Borghesi, S. V. Bulanov, G. Mourou, and T.
Tajima, Phys. Rev. Lett. 92, 175003 (2004).

\bibitem{Yin} L. Yin, B. J.  Albright, B. M. Hegelich and J. C.
Fernandez, Laser and Particle Beams 24(2), 291-298 (2006).

\bibitem{EsirkepovPRL96} T. Esirkepov, M. Yamagiwa, and T. Tajima, Phys. Rev. Lett. 96, 105001
(2006).

\bibitem{Henig} A. Henig, S. Steinke, M. Schn¨¹rer, T. Sokollik, R. H¡§orlein, D. Kiefer,
D. Jung, J. Schreiber, B. M. Hegelich, X. Q. Yan, T. Tajima, P. V.
Nickles, W. Sandner and D. Habs, arXiv:0908.4057v1 (2009).


\bibitem{YanxqPRL} X. Q. Yan, C. Lin, Z.M. Sheng, Z.Y. Guo, B.C. Liu, Y.R. Lu, J.X. Fang, and J.E.
Chen, Phys. Rev. Lett. 100, 135003 (2008).

\bibitem{unlimitedRPA} S. V. Bulanov, E. Yu. Echkina, T. Zh. Esirkepov, I. N. Inovenkov, M. Kando, F. Pegoraro, and G.
Korn, Phys. Rev. Lett. 104, 135003 (2010).

\bibitem{Yanxq} X. Q. Yan, T. Tajima, M. Hegelich, L. Yin and D.
Habs, Appl. Phys. B, 98£º711-721 (2010).

\bibitem{stableRPA} X. R. Hong, B. S. Xie, S. Zhang, H. C. Wu,
A. Aimidula, X. Y. Zhao, and M. P. Liu, Phys. Plasmas 17, 103107
(2010).

\bibitem{stableRPA1} T. P. Yu, A. Pukhov, G. Shvets, and M. Chen, Phys. Rev. Lett. 105, 065002
(2010).

\bibitem{Qiao2009}B. Qiao, M. Zepf, M. Borghesi, and M. Geissler, Phys. Rev. Lett. 102, 145002 (2009).

\bibitem{breaktime} A. Macchi, F. Cattani, T. V. Liseykina and F.
Cornolti, Phys. Rev. Lett. 94, 165003 (2005).

\bibitem{matlabfile} Supplement file.

\end{thebibliography}
\end{document}